\documentclass[10pt,conference]{IEEEtran}
\usepackage[T1]{fontenc}% optional T1 font encoding
\usepackage{amsmath}
\usepackage{bm}
\usepackage{amsmath}
\usepackage{cite}
\usepackage{graphicx}
\usepackage{amsfonts}
\usepackage{latexsym}
\usepackage{subfigure}
\usepackage{algorithm}
\usepackage{algorithmic}
\usepackage{color}
\usepackage{times}
\usepackage{epsfig, graphics}
\usepackage{bm}
\usepackage{subfigure}
\usepackage{graphicx,wrapfig,lipsum}
\usepackage{url}

\begin{document}

\newtheorem{theorem}{Theorem}
\newtheorem{lemma}{Lemma}
\newtheorem{conjecture}{Conjecture}
\newtheorem{corollary}{Corollary}
\newtheorem{definition}{Definition}
\newtheorem{scheme}{Scheme}

\newcommand{\argmax}{\arg\!\max}
\newcommand{\rev}[1]{{\color{red}#1}}
\newcommand{\pound}{\operatornamewithlimits{\gtrless}}
\IEEEoverridecommandlockouts

\title{Deep Learning for Launching and Mitigating Wireless Jamming
Attacks}
%\author{Tugba Erpek,~\IEEEmembership{Student Member,~IEEE,}
%Yalin E. Sagduyu,~\IEEEmembership{Senior Member,~IEEE,} and
%Yi~Shi,~\IEEEmembership{Senior Member,~IEEE}
\author{Tugba Erpek,
Yalin E. Sagduyu, and
Yi~Shi
\thanks{
T. Erpek is with Virginia Tech., Electrical and Computer Engineering, Arlington, VA, USA, and Intelligent Automation, Inc., Rockville, MD, USA; Email: terpek@vt.edu. Y. E. Sagduyu is with Intelligent Automation, Inc., Rockville, MD, USA; email: ysagduyu@i-a-i.com. Y. Shi is with Virginia Tech., Electrical and Computer Engineering, Blacksburg, VA, USA, and Intelligent Automation, Inc., Rockville, MD, USA; email: yshi@vt.edu.} \thanks{This effort is supported by the U.S. Army Research Office under contract W911NF-17-C-0090. The content of the information does not necessarily reflect the position or the policy of the U.S. Government, and no official endorsement should be inferred.} \thanks{A preliminary version of the material in this paper was partially presented at the Workshop on Promises and Challenges of Machine Learning in Communication Networks (ML4COM) at IEEE International Conference on Communications (ICC), 2018. }
}

\maketitle
\vspace{-1cm}
\begin{abstract}
An adversarial machine learning approach is introduced to launch jamming attacks on wireless communications and a defense strategy is presented. A cognitive transmitter uses a pre-trained classifier to predict the current channel status based on recent sensing results and decides whether to transmit or not, whereas a jammer collects channel status and ACKs to build a deep learning classifier that reliably predicts the next successful transmissions and effectively jams them.  This jamming approach is shown to reduce the transmitter's performance much more severely compared with random or sensing-based jamming. The deep learning classification scores are used by the jammer for power control subject to an average power constraint. Next, a generative adversarial network (GAN) is developed for the jammer to reduce the time to collect the training dataset by augmenting it with synthetic samples. As a defense scheme, the transmitter deliberately takes a small number of wrong actions in spectrum access (in form of a causative attack against the jammer) and therefore prevents the jammer from building a reliable classifier. The transmitter systematically selects when to take wrong actions and adapts the level of defense to mislead the jammer into making prediction errors and consequently increase its throughput.
\end{abstract}


\IEEEpeerreviewmaketitle

\section{Introduction}
Various detection, classification and prediction tasks are performed by \emph{cognitive radios} that use their task outcomes to learn from and adapt to their spectrum environment. These tasks can be effectively performed by \emph{machine learning} \cite{Clancy2007, Thilina2013} such as spectrum sensing, signal detection, channel estimation, and modulation classification.

Due to the broadcast nature of wireless communications, wireless medium is susceptible to adversaries that can learn and jam the ongoing transmissions. There are various ways of launching wireless \emph{jamming attacks} such as random and sensing-based jamming, and their impact can be detrimental to network performance \cite{XuMobihoc}. Different countermeasures against jamming have been developed using conventional methods such as frequency-hopping and transmit power control  \cite{JammingSurvey}. However, the vulnerabilities of cognitive radio systems using machine learning and potential mitigation techniques are not well understood yet. With the increasing use of machine learning in wireless communication systems, it is critical to understand the security implications of machine learning for cognitive radios. \emph{Adversarial machine learning} has emerged as a field that studies learning in the presence of an adversary and aims to enable safe adoption of machine learning to the emerging applications. In this paper, we leverage and advance techniques from adversarial machine learning to design jamming attacks and mitigation solutions.

We consider a communication system where both the transmitter and the jammer use machine learning to learn the spectrum and make their transmit decisions by adapting to the spectrum dynamics. We study a canonical wireless communication scenario with a transmitter, a receiver, a jammer, and another background traffic source. The transmitter senses the channel and transmits data to a receiver if the channel is found idle. While traditional algorithms for predicting idle channels may be as simple as comparing sensing results with some threshold (i.e., energy detector), more advanced techniques may be needed in a dynamic wireless environment with complex channel and transmitter characteristics. To determine the idle channel status, the transmitter first trains a supervised machine learning classifier using recent sensing results as the input and the status of the background transmitter  (on or off) as the labels. During normal operation, this pre-trained machine learning classifier classifies the channel as ``idle'' or ``busy'' by using the recent sensing results. We also discuss the generalization of this scenario to multiple transmitters and receivers.

In the context of adversarial machine learning, the jammer applies an \emph{exploratory attack} (also called \emph{inference attack}) to understand the outcome of the transmitter's operations as a preliminary step before jamming. Exploratory attacks have been considered in different data domains (such as image and text analysis) \cite{Ateniese,Tramer,Fredrikson,Shi17:HST, ShiMilcom}.
Note that in the general setting of an \emph{exploratory attack}, the jammer aims to build a classifier that is functionally equivalent to the target classifier under the attack, i.e., provides the same output as the target classifier for the same given input. While the attack considered in this paper shares this main idea, there is an important difference in the sense that the input and the output in classifiers are different for the target (i.e., the transmitter) and the adversary (i.e., the jammer) due to the following reasons:
\begin{enumerate}
  \item The input (sensing results) is different since the jammer and the transmitter are at different locations and are subject to different channels.
  \item The jammer does not need to distinguish idle channels or transmissions that are likely to fail. Instead, \emph{the jammer should predict whether there will be a successful transmission} (i.e., whether the transmitter will decide to transmit and the signal-to-interference-plus-noise ratio (SINR) will exceed a threshold at the receiver) so that it jams a transmission that would succeed if it was not jammed.
\end{enumerate}
The classifier built at the jammer is functionally equivalent to the one at the transmitter only in the sense that the jammer's classifier will decide to jam if and only if it predicts that there will be a successful transmission (in the absence of jamming) for the same instance. If the receiver successfully receives a packet, it sends an ACK as feedback to the transmitter; otherwise there is no feedback. During the learning period, the jammer senses the channel to determine whether there is an ACK or not, i.e., ACK signal plus noise vs. noise. The jammer needs to jam only if there will be an ACK. Thus, the jammer builds a \emph{deep learning} classifier (i.e., trains a deep neural network) with two labels (``ACK'' or ``no ACK'') by using the most recent sensing results (received signal strengths) as the features. The jammer has two objectives: minimize the misdetection probability (for effective jamming) and minimize the false alarm probability (to save energy or avoid being detected). Thus, the jammer only jams if it predicts there will be an ACK and aims to minimize the maximum of misdetection and false alarm probabilities of its prediction.

We show that the exploratory attack approach is very effective, i.e., for the scenario studied in numerical results, it reduces the transmitter's throughput from $0.766$ packet/slot to $0.050$ packet/slot. However, random jammer or sensing-based jammer that makes decisions based on instantaneous sensing results is not as effective since the transmitter can still sustain throughput of $0.383$ packet/slot against random jammer or $0.140$ packet/slot against sensing-based jammer (with the best sensing threshold). In addition, the success of sensing-based jammer strongly depends on the selection of the sensing threshold.

We also consider \emph{power control} at the jammer such that the jammer adjusts its transmit power subject to an average power constraint. In particular, the jammer selects its transmit power at any time slot as a function of the deep learning classification score that measures the likelihood of the jamming opportunity in that slot. We show that as the jammer's power budget is relaxed, its jamming success improves such that the performance (the throughput and the success ratio)  of the transmitter drops.

We show that the jamming performance drops when the jammer can only collect limited spectrum data (over limited time) for training purposes. To reduce the learning period before launching the attack, we develop the approach for the jammer to apply the \emph{generative adversarial network} (GAN) \cite{Goodfellow2014} that generates synthetic data based on a small number of real data samples in a short learning period and augments the training data with these synthetic data samples. The GAN consists of a generator and a discriminator playing a minimax game. The generator aims to generate realistic data (with labels), while the discriminator aims to distinguish data generated by the generator as real or synthetic. Our results show that the detection performance of the jammer that augments its training data with synthetic data is very close (within $0.19\%$ for misdetection and $3.14\%$ for false alarm) to the performance of the jammer that uses an increased number of real data samples (collected over a longer period of time) for training.

Other attacks such as \emph{evasion attacks} and \emph{causative or poisoning attacks} can be launched after exploratory attacks. These attacks have been extensively studied for other data domains such as computer vision. With evasion attacks \cite{Biggio,Kurakin}, the adversary attempts to fool the machine learning algorithm into making a wrong decision (e.g., fooling a security algorithm into accepting an adversary as legitimate). With causative or poisoning attacks \cite{Papernot2, Pi}, the adversary provides incorrect data to a machine learning-based application when it is re-trained in supervised learning. These attacks can be launched separately or combined, e.g., causative and evasion attacks can be launched by building upon the inference results of an exploratory attack \cite{ShiHST2018}.

Motivated by these attack schemes, we design a \emph{defense} scheme for the transmitter. The basic idea is to make the transmitter's behavior unpredictable, which can be enforced by the transmitter taking some deliberately wrong actions (i.e., transmitting on a busy channel or not transmitting on an idle channel) in some selected time slots. This corresponds to a \emph{causative} attack launched by the transmitter back at the jammer.
To maximize the impact of a small number of wrong actions, the transmitter uses the classification scores (an intermediate result) that are determined by the machine learning algorithm for spectrum sensing. Such a score is within $[0,1]$ and compared with a threshold to classify channels.
If this score is far away from the threshold (i.e., close to $0$ or $1$), the confidence of classification is high and the corresponding time instance should be selected to take the wrong action because it can more successfully deceive the jammer that aims to mimic the transmitter's behavior. A very small number of wrong decisions cannot fool the jammer. On the other hand, a large number of wrong decisions would prevent the transmitter from using the spectrum efficiently and reduce the performance significantly even in the absence of a jammer. Hence, there is a balance on how many wrong actions to take.

We show that by taking a small number of wrong actions on carefully selected time instances (based on its classifier's likelihood scores), the transmitter can fool the jammer into making a significant number of prediction errors. Thus, the transmitter's performance can be improved significantly from $0.050$ packet/slot to $0.216$ packet/slot. Note that random and sensing-based jamming cannot be mitigated by this defense scheme. We provide an \emph{adaptive defense} scheme that does not assume the knowledge of the jammer type and allows the transmitter to optimize its defense level on the fly based on its achieved throughput.

We start with a static scenario and then extend to two \emph{mobile} scenarios. In the first mobile scenario, we keep the distance between the jammer and the receiver fixed and let the jammer move on a circle around the receiver. In the second mobile scenario, we keep the distance between the background transmitter and the jammer fixed. We show that the jammer is more effective as it gets closer to the receiver (so the received power level at the receiver is higher) or to the background traffic source (so it can infer better when the transmitter will succeed next).

The rest of the paper is organized as follows. Section~\ref{sec:related_work} discusses the related work. Section~\ref{sec:scenario} describes the system model. Section~\ref{sec:transmitter} describes the transmitter's algorithm and shows the performance when there is no jamming. Section~\ref{sec:jammer} describes the jammer's algorithm and shows the performance under deep learning, sensing-based and random attacks, and introduces power control at the jammer. Section~\ref{sec:GAN} applies the GAN to reduce the learning period for the jammer. Section~\ref{sec:defense} presents the defense scheme and shows the attack mitigation performance. Section~\ref{sec:ext} includes the extension of network setting with mobile scenarios. Section~\ref{sec:conclusion} concludes the paper.

\section{Related Work}
\label{sec:related_work}

Machine learning has recently been applied to various problems in wireless communications \cite{Alsheikh2014, Chen2017}. In particular, deep learning has shown its potential to learn complex spectrum environments.  A convolutional neural network (CNN) was used in \cite{Lee2017} for spectrum sensing. A GAN was used in \cite{Kemal2018} to augment the training data for spectrum sensing. Channel estimation was performed in \cite{DeepOFDM} using a feedforward neural network (FNN) and modulation classification was performed in \cite{OShea2016} using a CNN.
	
A review on physical layer attacks of cognitive radio networks including sensing falsification, jamming and eavesdropping was provided in \cite{SecurePhy}. An adaptive, jamming-resistant spectrum access protocol was proposed in \cite{JamRes} for cognitive radio ad hoc networks,  where there are multiple channels that the secondary users can utilize. Jamming games between a cognitive user and a smart jammer was considered in \cite{UserCentric}, where they individually determine their transmit powers.

Game theory has been instrumental to analyze the conflicting interactions between transmitters and jammers. A game-theoretic framework to derive the optimal jamming attack and detection strategies was considered in \cite{GameJam} for wireless ad hoc networks. A jamming attack scenario when both the victim and the attacker are energy-constrained was considered in \cite{EnDep} through a multi-stage game formulation where the dynamics of full information is available at both sides. A stochastic game framework for an anti-jamming defense was  proposed in \cite{StochGame} assuming cognitive attackers can adapt their strategies to the time-varying spectrum environment and the strategies of secondary users.

A deep Q-network algorithm was used in \cite{TwoDim} for secondary users to decide whether to leave an area of heavy jamming or choose a frequency-hopping pattern to defeat smart jammers. Cumulative sum method was used in \cite{IDS} as a statistical intrusion detection mechanism at the secondary user. The throughput optimization problem for the secondary users under jamming attacks was formulated in \cite{EnergyHarvestingCN}  as a Markov Decision Process and a learning algorithm was proposed for the secondary user to find an optimal transmission policy. The jammer does not typically use machine learning techniques in these works. A spectrum data poisoning attack was considered in \cite{Yi2018_2} by employing deep learning to manipulate the training data collected during spectrum sensing and fool a transmitter into making wrong transmission decisions. In this paper, we focus on jamming the data transmissions, apply deep neural networks to learn the subtle traffic, channel, and interference characteristics leading to the transmission success, and build the jamming strategy accordingly to maximize its impact on the transmitter's communication performance. Deep learning is used at both the transmitter and the jammer to make transmission and jamming decisions, respectively. A preliminary version of the material in this paper was partially presented in \cite{Yi2018} under limited network, channel, and jammer models.

\section{System Model}
\label{sec:scenario}

\begin{figure}
	\centering
	\includegraphics[width=1.0\columnwidth]{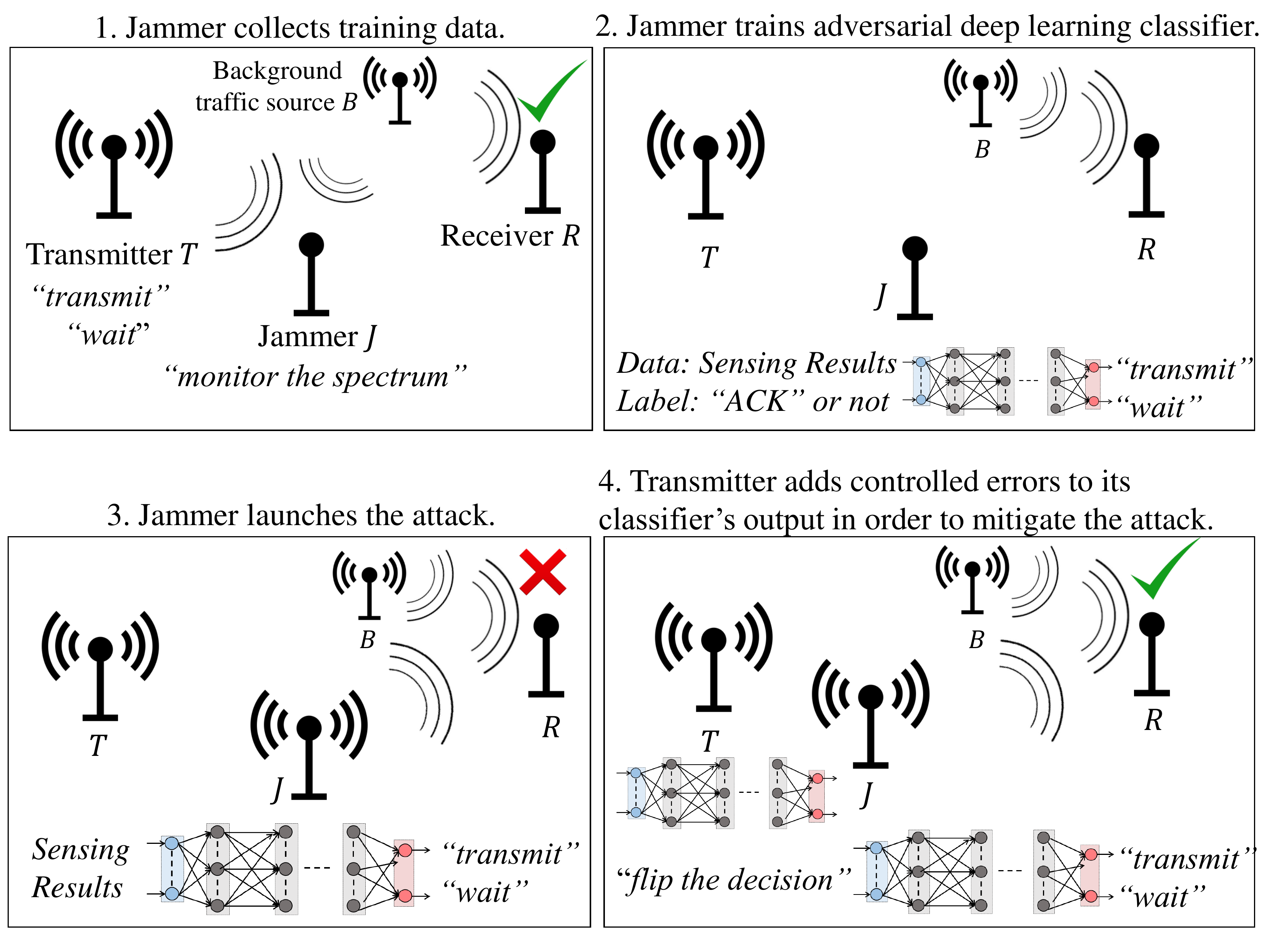}
	\caption{The scenario and the steps to launch and mitigate the attack.}\label{fig:scenario}
\end{figure}

Fig.~\ref{fig:scenario} shows the system model. We consider a wireless communication scenario with one transmitter $T$, one receiver $R$, and one jammer $J$. This setting is instrumental in studying the fundamentals of jamming and defense strategies in wireless access \cite{Sagduyu2011}. There is a background traffic source $B$, whose transmission behavior is not known by either $T$ or $J$.
As will be discussed in Section~\ref{sec:ext}, the developed algorithms can be easily extended to multiple mobile transmitters and receivers, while a single jammer alone can jam nodes within its transmission range.

After $T$ builds its deep learning classifier (to make transmission decisions), $J$ launches its attack using the following three main steps (see Fig.~\ref{fig:scenario}):
\begin{enumerate}
	\item $J$ senses the spectrum and collects training data.
	\item $J$ trains its adversarial deep learning classifier (that predicts whether there will be a successful transmission).
	\item $J$ jams the predicted successful transmissions.
\end{enumerate}
Fig.~\ref{fig:scenario} also shows that $T$ defends against this attack by adding controlled errors to its decisions to mislead the training process of $J$.

\subsection{Operation Modes}

There is a single channel and time is divided in slots. The channel busy/idle status is set up by introducing the background traffic source as follows.
\begin{itemize}
\item \emph{Background traffic source's operation}: We assume random packet arrivals at the background traffic source $B$ according to the Bernoulli process with rate $\lambda$ (packet/slot). If $B$ is not transmitting, it becomes active with certain probability when its queue is not empty. Once activated, it will keep transmitting until its queue becomes empty.
    Thus, there may be a continuous period of busy slots. The length of such a busy period depends on the number of previous idle slots.
    Hence, channel busy/idle states are correlated over time. Therefore, both $T$ and $J$ need to observe the past channel status over several time slots to predict the current channel status.
\end{itemize}

The operation models for $T$, $R$ and $J$ are given as follows.
\begin{itemize}
\item \emph{Transmitter's operation}: Since there may be transmissions from $B$ at any time, the channel status may be busy even when $T$ and $J$ do not transmit. In each slot, the short initial period of time is allocated for $T$ to sense the channel, run its spectrum sensing algorithm, and detect the channel (idle/busy) status. The length of a slot should be no more than the time between changes in channel (idle/busy) status. If the channel is detected as idle, $T$ can transmit data to $R$ with power $P_T$ in this time slot.\footnote{We consider fixed transmit power since even if a transmitter can perform power control, it will always use the maximum power (namely, fixed power, $P_{max}$ allowed by its hardware) such that the success probability at the receiver can be maximized.} Otherwise, $T$ remains idle. Without loss of generality, we assume that $T$ transmits only one packet in a time slot.
\item \emph{Jammer's operation:} $J$ also senses the spectrum and predicts whether there will be a successful transmission (with feedback ACK), or not (without a feedback) in a time slot. If $J$ predicts that there will be a successful transmission, it jams with power $P_J$ in this time slot.\footnote{We initially focus on a fixed jamming power. Extension to adaptive jamming power is discussed in Section~\ref{subsec:pj}.} Otherwise, $J$ remains idle.
\item \emph{Receiver's operation}: If a transmission is not jammed, the SINR at $R$ is $\frac{g_{TR} P_T}{N_0 + I_R}$ if channel is busy, or $\frac{g_{TR} P_T}{N_0}$ if channel is idle, where $I_R$ is the interference from some unobserved transmitters to $R$, $N_0$ is a Gaussian noise with power normalized as one, and $g_{TR}$ is the channel gain between $T$ and $R$.
If a transmission is jammed, the SINR at $R$ is reduced to $\frac{g_{TR} P_T}{N_0 + I_R + g_{JR} P_J}$ (if channel is busy) or $\frac{g_{TR} P_T}{N_0 + g_{JR} P_J}$ (if channel is idle), where $g_{JR}$ is the channel gain between $J$ and $R$. It is assumed that the signal strength diminishes proportionally to $1/d^2$, where $d$ is the distance between a transmitter and a receiver. Log-normal shadowing is used as the shadowing model. A transmission is successful if the SINR at $R$ is larger than some threshold $\beta$. The short ending period of a time slot is allocated for $R$ to send feedback (ACK) to $T$.
\end{itemize}
This general operation does not specify a particular algorithm for $T$ to make transmission decisions or for $J$ to jam the channel.
Note that $J$ does not jam all time slots, although doing so can maximize the success of jamming.
There are two reasons to jam only the selected time slots.
First, $J$ will be easily detected if it is jamming in all time slots due to the high false alarm rate.
Second, $J$ may have power budget in terms of the average jamming power, and thus cannot jam all time slots.
In Section~\ref{subsec:pj}, we will present results for a jammer with limited power budget.

\subsection{Classifiers of Transmitter and Jammer}
We consider the case that both $T$ and $J$ apply some machine learning algorithms (unknown to each other) to make their transmission decisions. Denote sensing results (noise power or noise plus interference power) at time $t$ as $s_T(t)$ and $s_J(t)$ for $T$ and $J$, respectively.
Note that due to different locations of $T$ and $J$ and different channel effects, their sensing results may be different in general, i.e., $s_T(t) \neq s_J(t)$ at any time $t$. The classifiers used by $T$ and $J$ for transmission and jamming decisions have the following properties.
\begin{itemize}
  \item \emph{Transmitter's classifier}: $T$ has a classifier $C_T$ that is pre-trained by some machine learning algorithm and identifies the current time slot $t$ as idle or busy based on recent $K_T$ sensing results $(s_T(t-K_T+1), \cdots, s_T(t-1), s_T(t))$.
      $T$ uses
  \begin{equation} \label{eq:featureT}
  \bm{x}_T(t) = (s_T(t-K_T+1), \cdots, s_T(t-1), s_T(t))
  \end{equation}
   as its features in time slot $t$ and
  \begin{equation} \label{eq:labelT}
   y_T(t) = \{``\text{idle}", ``\text{busy}"\}
   \end{equation}
   as its labels  in time slot $t$ to build its training data $\{(\bm{x}_T(t),y_T(t))\}_t$. Here, ``\text{idle}'' or ``\text{busy}'' means that the channel is idle or busy, respectively.
      Note that $T$ uses multiple sensing results as its features since features should be able to capture time correlation and help achieve a high sensing accuracy in a short period of time.
      Then classifier $C_T: \bm{x}_T(t) \mapsto y_T(t)$ defines the mapping from  sensing results to sensing decision and consequently to transmission decision in time slot $t$.
  \item \emph{Jammers's classifier}: $J$ does not know the classifier $C_T$ and needs to build another classifier $C_J$ itself by training a deep learning classifier, which predicts whether there will be a successful transmission, or not, in time slot $t$ based on recent $K_J$ sensing results $(s_J(t-K_J+1), \cdots, s_J(t-1), s_J(t))$. $J$ uses
  \begin{equation} \label{eq:featureJ}
  \bm{x}_J(t) = (s_J(t-K_J+1), \cdots, s_J(t-1), s_J(t))
  \end{equation}
   as its features  in time slot $t$ and the presence and absence of ACK after transmission, namely
   \begin{equation} \label{eq:labelJ}
   y_J(t) = \{``\text{ACK}", ``\text{no ACK}"\}
   \end{equation}
     as its labels in time slot $t$ to build its training data. Here, ``\text{ACK}'' or ``\text{no ACK}'' means that there is an ACK following a transmission, or not, respectively. Then classifier $C_J: \bm{x}_J(t) \mapsto y_J(t)$ defines the mapping from  sensing results to  prediction of successful transmission and consequently to jamming decision in time slot $t$.
\end{itemize}
The deep neural network is a universal approximator and is expected to match and exceed the performance of any rule-based scheme (such as an energy detector). Therefore, we assume transmitter and jammer use \emph{deep neural networks} as classifiers $C_T$ and $C_J$, respectively, each with different hyperparameters selected to optimize their individual performance.

\subsection{Performance Measures}
To evaluate the accuracy of a classifier, we define error probabilities on misdetection $e_{MD}$ and false alarm $e_{FA}$ for $T$ and $J$.
\begin{itemize}
	\item \emph{Misdetection} for $T$: A time slot is idle, but $T$ predicts it as busy.
	\item \emph{False alarm} for $T$: A time slot is busy, but $T$ predicts it as idle.

	\item \emph{Misdetection} for $J$: $T$'s transmission is successful, but $J$ predicts there will not be an ACK.
	\item \emph{False alarm} for $J$: $T$ does not transmit or $T$'s transmission fails (even without jamming), but $J$ predicts that there will be an ACK.
\end{itemize}
$T$ and $J$ individually optimize the hyperparameters of their own deep neural networks (namely their classifiers) to minimize their own value of $\max\{ e_{MD}, e_{FA}\}$.
This objective ensures a small error for each class.

In addition, we also measure $T$'s performance by throughput and success ratio.
\begin{itemize}
	\item \emph{Throughput}: The number of received packets at $R$ during a period divided by the number of time slots in this period.
	\item \emph{Success ratio}: The percentage of successful transmissions  by $T$ during a period over all transmissions during this period.
\end{itemize}

\section{Transmitter's Operation}
\label{sec:transmitter}
$T$ applies a deep learning (deep neural network) based classifier to determine the channel status. Note that $T$ could also use a simpler machine learning algorithm at the expense of potential performance loss. $T$ senses the channel and records the most recent $K_T$ received signal strength indicator (RSSI) results. In time slot $t$, $T$ uses these sensing results to build the feature set $\bm{x}_T(t)$  given in (\ref{eq:featureT}). For numerical results, we assume $K_T = 10$. In time slot $t$, $T$ uses the current channel busy/idle status as its label $L_T(t)$.
Each result is either a Gaussian noise $N_0$ with normalized unit power (when the channel is idle) or noise plus the transmit power from the unobserved transmitter received at $T$, i.e., $N_0 + I_T$ (when the channel is busy), where $I_T$ is the interference received at $T$.

After observing a certain period of time, $T$ collects a number of samples as training data to train the deep learning classifier $C_T$ that returns labels ``idle'' and ``busy''. For that purpose, $1000$ samples are collected by $T$ and split by half to build its training and test data. $T$ trains an FNN as the deep learning classifier $C_T$. The Microsoft Cognitive Toolkit (CNTK) \cite{CNTK} is used to train the FNN. $T$ optimizes the hyperparameters of the deep neural network to minimize $\max\{ e_{MD}, e_{FA}\}$. When the arrival rate $\lambda$ for $B$ is $0.2$ (packet/slot), the optimized hyperparameters of $C_T$ are found as follows:
\begin{itemize}
	\item The neural network consists of one hidden layer with $100$ neurons.
	\item The cross entropy loss function (given in (\ref{eq:cost})) is minimized to train the neural network with backpropagation algorithm \cite{Backprop}.
\begin{equation}
\begin{aligned}
\label{eq:cost}
C(\bm{\theta}) = - \sum_i & \left[\bm{y}_T\right]_i \log\left(\left([a^L(\bm{x}_T)\right]_i\right) + \\
& \left(1-\left[\bm{y}_T\right]_i\right) \log\left(1-[a^L\left(\bm{x}_T\right)]_i\right),
\end{aligned}
\end{equation}
where $\bm{\theta}$ is the set of the neural network parameters,
$\bm{x}_T$ is the training data vector, $\bm{y}_T$ is the corresponding label vector, and $a^L\left(\bm{x}_T\right)$ is the output of the neural network at the last layer $L$.
	\item The output layer uses softmax activation such that for input $\bm{z}$, the $k$th entry of the activation function output is given by $[\sigma(\bm{z})]_k = \frac{e^{\bm{z}_k}}{\sum_j e^{\bm{z}_j}}$.
	\item The hidden layers are activated using the sigmoid function  such that for input $\bm{z}$, the $k$th entry of the activation function output is given by $[\sigma(\bm{z})]_k = \frac{1}{1+e^{-\bm{z}_k}}$.
	\item All weights and biases are initialized to random values in $[-1.0,1.0]$.
	\item The input values are unit normalized in the first training pass.
	\item The minibatch size is  $25$.
	\item The momentum coefficient to update the gradient is $0.9$.
	\item The number of epochs per time slot is $10$.
\end{itemize}
After training the FNN with these hyperparameters, we run $T$'s classifier over $500$ time slots to evaluate its performance. We set $P_B = P_T = 1000 N_0$, and $\beta = 3$. The positions of the $T$, $R$ and $B$ are fixed at locations $(0,0), (10,0)$, and $(0,10)$, respectively. We have $d_{TR} = 10$ and $d_{TB} = 10$. We extend the model to a mobile network in Section \ref{sec:ext}. For these scenario parameters, $T$ can build a very good classifier (in fact we have $e_{MD}=e_{FA}=0$). $T$ makes $400$ transmissions and $383$ of them are successful. Note that $17$ transmissions on idle channels fail due to random channel conditions. Thus, the throughput is $383/500=0.766$ packet/slot and the success ratio is $383/400=95.75\%$. In Section \ref{sec:jammer}, we will show how adversarial deep learning-based jammer can reduce this performance significantly.

\section{Jammer's Operation}
\label{sec:jammer}

\subsection{Adversarial Deep Learning-based Jammer}
\label{subsec:dl-jammer}

It is viable to infer a machine learning (including deep learning) classifier via the \emph{exploratory attack} that has been applied to text classification in \cite{Shi17:HST} and to image classification in \cite{ShiMilcom}. In these previous works, the basic idea was to call the target classifier and obtain labels of a number of samples and then train a functionally equivalent classifier using deep learning. Two classifiers are functionally equivalent if they provide the same labels for the same sample. However, this approach cannot be applied to the setting in this paper.
Due to different locations of $T$ and $J$, random channel gain and random noise, the sensing results at $T$ and $J$ will be different. That is, when the channel is idle, both $T$ and $J$ sense a Gaussian noise $N_0$ but the value may be different due to different realizations. When the channel is busy, $T$ will sense $N_0 + I_T$ and $J$ will sense $N_0 + I_J$. Thus, in addition to different realizations of $N_0$, the values of $I_T$ and $I_J$ are different due to different channel gains to $T$ and $J$, as well as their different realizations. Thus, even if $J$ has a functionally equivalent classifier (e.g., $T$'s classifier), $J$ cannot use it to obtain the same channel status as the one predicted by $T$ due to different sensing results (or features computed for deep learning). Moreover, $J$ does not aim to predict whether the channel is idle or busy. Instead, its goal is to predict whether there will be a successful transmission by $T$, or not. These are different goals since a successful transmission depends on the SINR and therefore it depends not only on the presence of background signal but also on channel gain and noise. In addition, the perceptions of sensed signals are different because of different locations and channel instances.

There are four cases for the channel status and $T$'s behavior:
\begin{enumerate}
\item channel is idle and $T$ is transmitting,
\item channel is busy and $T$ is not transmitting,
\item channel is idle and $T$ is not transmitting, and
\item channel is busy and $T$ is transmitting.
\end{enumerate}
Ideally, the last two cases should be rare cases, since they refer to wrong sensing decisions by $T$.
$J$ needs to detect the presence of an ACK message transmission when $T$'s transmission is successful. Note that $J$ does not attempt to decode ACK messages.\footnote{ACK messages are typically distinct from data messages in the sense that they are short and follow the data transmission with some time lag. To ensure ACK can be reliably received, it is usually coded such that the required SNR to receive the ACK is much smaller than that for data packet. Hence, we assume that $J$ can reliably detect the presence of ACK message transmissions and distinguish them from data transmissions.} Then $J$ can use the most recent $10$ sensing results as features (i.e., $L = 10$) and the current feedback (ACK or no ACK) as a label to build one sample. $J$ aims to jam successful transmissions (with received ACK feedback) only. Thus, $J$ defines two labels as ``ACK'' (i.e., ``a successful transmission'') and ``no ACK''  (i.e., ``no successful transmission''). Thus, the labels at $J$ are also different from $T$. In summary, we contrast the classifiers for $T$ and $J$ as follows.
For $T$'s classifier,
\begin{itemize}
  \item the features for deep learning are $T$'s sensing results and
  \item the predicted labels are ``idle'' and ``busy'',
\end{itemize}
while for $J$'s classifier,
\begin{itemize}
  \item the features are $J$'s sensing results and
  \item the predicted labels are ``ACK'' and ``no ACK''.
\end{itemize}
After observing a certain number of time slots, $J$ collects a number of samples to be used as training data and trains a deep learning classifier. Once the classifier is built, $J$ uses it to predict the status in any future time slot, i.e., whether there will be a successful transmission, or not. If yes, $J$ transmits in this slot. Note that when $J$ launches its attack, it does not need to collect ACKs anymore.

$J$ trains an FNN as the deep learning classifier $C_J$. For that purpose, $1000$ samples are collected by $J$ and split by half to build its training and test data. $J$ optimizes the hyperparameters to minimize $\max\{ e_{MD}, e_{FA}\}$. The training time (including hyperparameter optimization) is $67$ seconds and the test time per sample is $0.024$ milliseconds. The optimized hyperparameters of the $C_J$ are found as follows.
\begin{itemize}
	\item The neural network consists of two hidden layers with $50$ neurons.
	\item The cross entropy loss function is used to train the deep neural network with backpropagation algorithm.
	\item The output layer uses softmax activation.
	\item The hidden layers are activated using the hyperbolic tangent (Tanh) function such that for input $\bm{z}$, the $k$th entry of the activation function output is given by $[\sigma(\bm{z})]_k = \frac{e^{\bm{z}_k}-e^{-\bm{z}_k}}{e^{\bm{z}_k}+e^{-\bm{z}_k}}$.
	\item All weights and biases are initialized to random values in $[-1.0,1.0]$.
	\item The input values are unit normalized in the first training pass.
	\item The minibatch size is  $25$.
	\item The momentum coefficient to update the gradient is $0.9$.
	\item The number of epochs per time slot is $10$.
\end{itemize}

After training the deep neural network with these hyperparameters, we run classifiers of $J$ and $T$ over $500$ time slots to evaluate the attack performance. The positions of $T$, $R$ and $B$, and their transmit powers are given in Section~\ref{sec:transmitter}. The position of $J$ is fixed at location $(10,10)$ and its jamming power is $P_J = 1000 N_0$. Thus, we have $d_{JR} = 10$ and $d_{JT} = 10 \sqrt{2}$. In these time slots, if there is no attack (i.e., no jamming), $T$ will have $383$ successful transmissions. Under $J$'s attack, the number of misdetections is $16$, i.e., misdetection probability is $e_{MD}=16/383=4.18\%$ (majority of successful transmissions are jammed), and the number of false alarms is $17$, i.e., false alarm probability is $e_{FA}=17/(500-383)=14.53\%$.
The impact of this attack is significant. There are only $25$ successful transmissions among $400$ transmissions. Thus, the throughput of $T$ is reduced from $0.766$ packet/slot to $25/500=0.05$ packet/slot and the success ratio of $T$ is reduced from $95.75\%$ to $25/400=6.25\%$.

\subsection{Sensing-based and Random Jamming as Benchmark Schemes}
\label{sec:sensing-based}

For comparison purposes, we consider two alternative jamming schemes for $J$.
\begin{enumerate}
\item \emph{Sensing-based jamming}: $J$ jams the channel if its received power (noise or noise plus signal) during spectrum sensing in the current slot is greater than a threshold $\tau$. The performance of a sensing-based jammer is highly dependent on the selected threshold $\tau$. If $\tau$ is selected too low, the number of false alarms increases. Then $J$ consumes more power than required and this is not desired if it has a power constraint or objective. If $\tau$ is selected too high, then the number of misdetections increases. This threshold is usually given as a fixed value since adjusting the threshold optimally for a given false alarm and misdetection rate is difficult without using past sensing results and tracking the ACK messages from the receiver as a feedback.
We evaluated the performance of the sensing-based jammer with different $\tau$ values. Then we select $\tau$ that minimizes $\max\{ e_{MD}, e_{FA}\}$ and use it to compute the throughput and the success ratio of the transmitter in the presence of sensing-based jammer. This approach allows for a fair comparison between the performance of the sensing-based and adversarial deep learning-based jammer. Table \ref{table:sensingBasedSelection} shows the misdetection and false alarm probabilities for $\tau$ values from $1$ to $9$. $\tau = 1$ is the lowest $\tau$ that  makes misdetection rate 0\% based on the observed RSSI levels. However, it provides a high false alarm rate. $\tau = 3.4$ is determined to be the best threshold that minimizes $\max\{ e_{MD}, e_{FA}\}$.

\item \emph{Random jamming}:  $J$ jams the channel in some randomly selected instances. Such an attack scheme does not require $J$ to learn the channel's busy/idle status. As in deep learning-based and sensing-based jammers, jamming probability $p_J$ is selected to minimize $\max\{ e_{MD}, e_{FA}\}$. The jammer and transmitter actions are independent from each other, therefore $e_{MD} = 1-p_J$ and $e_{FA} = p_J$. As a result, $p_J$ is selected as $0.5$ with $\max\{ e_{MD}, e_{FA}\} = 50\%$. The impact of this attack is not as significant as deep learning based-jamming. The throughput can only be reduced from $0.766$ packet/slot to $0.383$ packet/slot and the success ratio can only be reduced from $95.75\%$ to $47.88\%$.
\end{enumerate}

The results are summarized in Table~\ref{table:summary} with using the best parameters (threshold or probability) in all three jamming approaches. Both sensing-based jamming and random jamming are not as effective as deep learning-based jamming since random jamming does not utilize any sensing result while sensing-based jamming only uses the sensing result in the current time slot.
On the other hand, deep learning-based jamming considers sensing results in recent time slots and generates a score for the likelihood of an ACK. In addition, deep learning tunes a threshold for these scores to minimize the classification errors. Thus, the deep learning-based jamming performs better than the other two attack schemes.

\begin{table}
\caption{Determining the sensing threshold for sensing-based jammer.}
\centering
{\small
\begin{tabular}{c|c|c|c}
Threshold & Misdetection & False alarm & $\max\{ e_{MD}, e_{FA}\}$ \\
 & prob. ($e_{MD}$) & prob. ($e_{FA}$) & \\ \hline \hline
$1$ & $0.0\%$ & $14.2\%$ & $14.2\%$ \\ \hline
$2$ & $0.6\%$ & $14.0\%$ & $14.0\%$ \\ \hline
$3$ & $8.8\%$ & $13.4\%$ & $13.4\%$ \\ \hline
$3.4$ & $12.8\%$ & $12.6\%$ & $12.8\%$ \\
(Best Thr.) & & & \\ \hline
$4$ & $19.2\%$ & $10.2\%$ & $19.2\%$ \\ \hline
$5$ & $30.0\%$ & $7.0\%$ & $30.0\%$ \\ \hline
$6$ & $44.2\%$ & $5.0\%$ & $44.2\%$ \\ \hline
$7$ & $52.2\%$ & $3.6\%$ & $52.2\%$ \\ \hline
$8$ & $56.8\%$ & $2.6\%$ & $56.8\%$ \\ \hline
$9$ & $60.6\%$ & $1.4\%$ & $60.6\%$
\end{tabular}
}
\label{table:sensingBasedSelection}
\end{table}

\begin{table}
\caption{Effect of different attack types on the transmitter's performance.}
\centering
{\small
\begin{tabular}{c|c|c}
Attack type & Throughput & Success ratio \\ \hline \hline
No attack & $0.766$ & $95.75\%$ \\ \hline
Adversarial deep learning & $0.050$ & $6.25\%$ \\ \hline
Sensing-based attack ($\tau = 3.4$) & $0.140$ & $16.99\%$ \\ \hline
Random attack & $0.383$ & $47.88\%$
\end{tabular}
}
\label{table:summary}
\end{table}

\subsection{Jamming Power Control with an Average Power Constraint}
\label{subsec:pj}

In this section, we consider the case that $J$ has some power budget in terms of the average jamming power $P_{avg}$ per slot.
That is, we have
\begin{equation}
\frac{1}{T} \sum_{t=1}^T p_t \le P_{avg}
\end{equation}
for large $T$, where $p_t$ is the jamming power in slot $t$. $J$ applies power control by adjusting $p_t$ in $[P_{min}, P_{max}]$. The deep learning classifier not only provides the label for each slot, but also a score for this classification. This score $s$ is compared with a threshold $S$ and if $s$ in a slot satisfies $s \le S$, it is classified as a slot with ACK. Then $s$ also represents the confidence of classification, i.e., the confidence of classification is high if $s$ is small.
Thus, a solution of power control should be a decreasing function $p(s)$ of $s$ in $[0, S]$ and $p(s)$ value should be either within $[P_{min}, P_{max}]$ or zero (i.e., no transmission).

We develop a piecewise linear function $p(s)$ as follows.
\begin{enumerate}
\item Denote the average power required by function $p(s)=P_{min}$ as $P_1$. If $P_1 \ge P_{avg}$,  $J$ may not jam all slots with $s \le S$ even using the minimum jamming power. Thus,  $J$ should not jam some slots with low classification confidence, i.e., with $s$ close to $S$. We need to find a suitable $c \in [0, S]$ such that the following $p(s)$ meets the power budget: $p(s)=P_{min}$ if $s \le c$, and $p(s) = 0$ otherwise. This $c$ value can be determined by the training data used to train $J$'s classifier.

\item Denote the average power required by function $p(s)=P_{max}$ as $P_2$. If $P_2 \le P_{avg}$, $J$ can jam all slots with $s \le S$ even using the maximum jamming power. Thus, we have $p(s)=P_{max}$. The scenario in Section~\ref{subsec:dl-jammer} can be regarded as this special case when we set $P_{avg} = P_J$.

\item In the above two cases, $P_{avg}$ is either very small or very large. As a consequence, jamming power is fixed as either $P_{min}$ or $P_{max}$. For the other cases, denote the average power required by function $p(s)=P_{max} - c_1 s$ as $P_3$, where $c_1 = \frac{P_{max}-P_{min}}{S}$. For this linear function, we have $p(0) = P_{max}$ and $p(S) = P_{min}$.
\begin{enumerate}
\item If $P_3 = P_{avg}$, we use $p(s)=P_{max} - c_1  s$.
\item If $P_3 > P_{avg}$, $J$ needs to use a smaller jamming power for each slot with $s \le S$. In particular, we consider $p(s)= \max\{P_{max} - c  s, P_{min}\}$ for some $c > c_1$ to meet the power budget. This $c$ value can be determined by the training data used to train $J$'s classifier.
\item If $P_3 < P_{avg}$, $J$ can use a larger jamming power for each slot with $s \le S$. In particular, we consider $p(s)= \min\{P_{min} + c (S-s), P_{max}\}$ for some $c > c_1$ to meet the power budget. This $c$ value can be determined by the training data used to train $J$'s classifier.
\end{enumerate}
\end{enumerate}
We assume that the distribution of score $s$ in the training data is the same as the distribution of $s$ in the test data in the above analysis and algorithm design. In reality, these two distributions are similar but may not be exactly the same. As a result, we may observe that $\frac{1}{T} \sum_{t=1}^T p_t > P_{avg}$ for some large $T$. Instead, the achieved power budget $P \left( \frac{1}{T} \sum_{t=1}^T p_t \le P_{avg} \right)$ is close to $1$
for large $T$.

\begin{table}
\caption{The effect of jammer's power budget on the transmitter's performance.}
\centering
\small
\begin{tabular}{c|c|c}
$P_{avg}$ & Throughput (packet/slot) & Success ratio \\ \hline \hline
$0$ & $0.766$ & $95.75\%$ \\ \hline
$0.2 \cdot P_{max}$ & $0.428$ & $53.50\%$ \\ \hline
$0.4 \cdot P_{max}$ & $0.170$ & $21.25\%$ \\ \hline
$0.6 \cdot  P_{max}$ & $0.094$ & $11.75\%$ \\ \hline
$0.8 \cdot  P_{max}$ & $0.066$ & $8.25\%$ \\ \hline
$P_{max}$ & $0.050$ & $6.25\%$
\end{tabular}
\label{table:summary2}
\end{table}

Table~\ref{table:summary2} shows results for different $P_{avg}$ values, where we assume $P_{min} = 500 N_0$ and $P_{max} = 1000 N_0$.
Note that the two extreme cases $P_{avg}=0$ and $P_{avg}=P_{max}$ correspond to the case of no attack and the case of no power budget, respectively.
Thus, the performance are the same as those in the first two rows of Table~\ref{table:summary}.
For other cases, results in Table~\ref{table:summary2} show that with a larger power budget, $J$ can successfully jam more transmissions and thus $T$'s performance (throughput and success ratio) becomes worse.

\section{Jamming Attack with Limited Training Data Collected}
\label{sec:GAN}
\begin{figure}
  \centering
  \includegraphics[width=0.7\columnwidth]{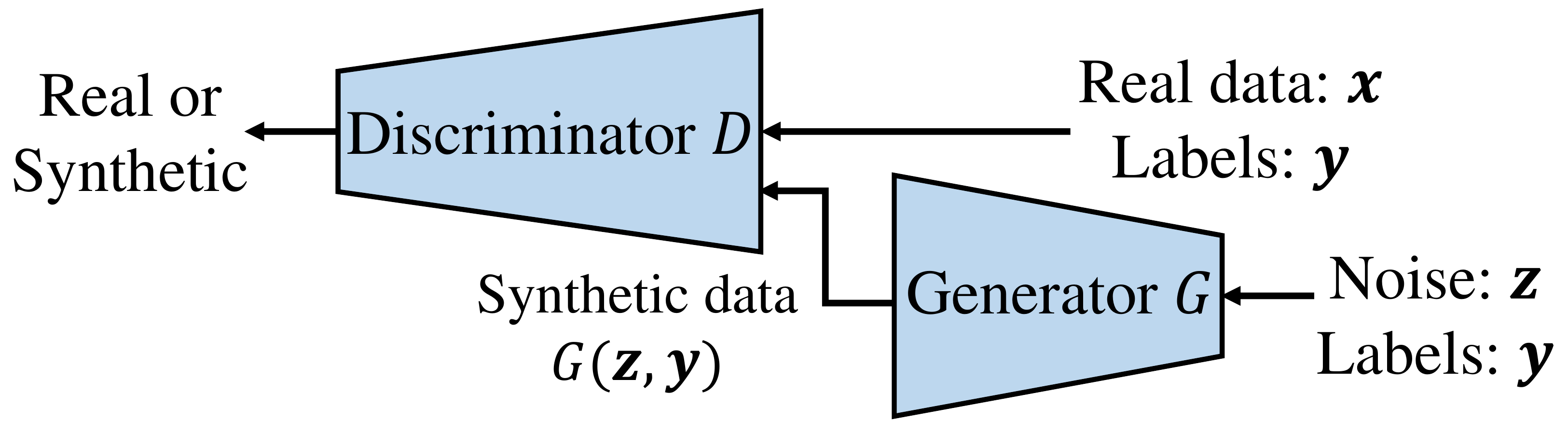}
  \caption{Conditional GAN for training data augmentation.}\label{fig:cgan}
\end{figure}

The attack results presented in Sec. \ref{sec:jammer} are obtained by assuming that $J$ collects $500$ samples to build its classifier $C_J$ with small errors and $500$ samples to test it. However, this may require a long learning period before launching the attack. It is important to shorten the initial learning period of $J$ and reduce the response time of $J$ to start jamming. To reduce this learning period, $J$ applies the GAN \cite{Goodfellow2014} approach to generate synthetic data based on real data in a short learning period. Then it uses these synthetic data samples to augment the training data. Synthetic data has been widely used in machine learning applications to extend the training datasets. Conventionally, image processing applications are known to benefit from additional data generated by scaling and rotating the existing images. Recently, GAN is used to generate synthetic images to augment training data for computer vision, text, and cyber applications \cite{Shrivastava2017, YiISSPIT2018, bookchapter2018}. In this paper, we use the GAN to generate synthetic spectrum data. In the GAN, there are two neural networks, namely the {\em generator\/} $G$ and the {\em discriminator\/} $D$, playing a minimax game formulated as:
\begin{align} \min_{G} \max_{D} \mathbb{E}_{\bm{x} \sim p_{data}} [\log(D(\bm{x}))] - \mathbb{E}_{\bm{z} \sim p_{\bm{z}}} [\log(1 - D(G(\bm{z})))],
\label{eqn:gan_obj}
\end{align}
where $\bm{z}$ is a noise input to generator $G$ with a model distribution of $p_{\bm{z}}$ and $G(\bm{z})$ is the generator output. Input data $\bm{x}$ has distribution $p_{data}$ and discriminator $D$ distinguishes between the real and generated samples. Both $G$ and $D$ are trained using the backpropagation algorithm. However, when $G$ is trained with the objective in (\ref{eqn:gan_obj}), the gradients
of $G$ rapidly vanish that makes the training of GAN very difficult. To address the vanishing gradient problem,
it is proposed to use
\begin{align} \max_{G} \mathbb{E}_{\bm{z} \sim p_{\bm{z}}} [\log(1 - D(G(\bm{z})))]
\label{eqn:gan_subs}
\end{align}
as the objective function at $G$ \cite{Goodfellow2014}.
In the first step, $D$ is trained to distinguish between the real and synthetic data. In the second step, $G$ takes random noise as input and maximizes the probability of $D$ making a mistake by creating samples that resemble the real data. The original GAN implementation does not include labels. Conditional GAN extends the GAN concept such that the generator can generate synthetic data samples with labels \cite{Mirza2014}. The objective function in conditional GAN is similar to the one of a regular GAN, but the terms $D(\bm{x})$ and $G(\bm{z})$ are replaced by $D(\bm{x},\bm{y})$ and $G(\bm{z},\bm{y})$, respectively, to accommodate the labels $\bm{y}$ as conditions.
The conditional GAN used by the jammer is shown in Fig. \ref{fig:cgan}.

\begin{figure}
  \centering
  \includegraphics[width=\columnwidth, trim={2cm 8cm 2cm 8cm}, clip]{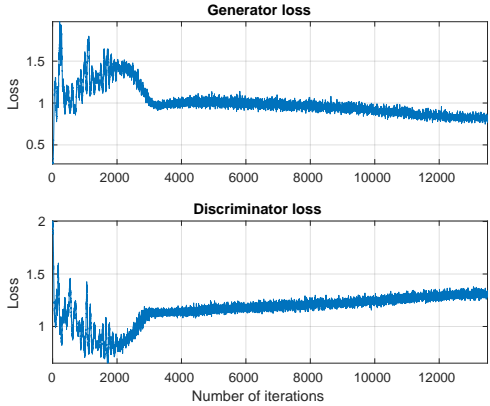}
  \caption{Discriminator and generator losses during training.}\label{fig:dgloss}
\end{figure}

We implemented the conditional GAN in TensorFlow \cite{Tensorflow} by using the FNNs with two hidden layers for both $G$ and $D$. Leaky ReLu $\sigma(x) = \max(x, \alpha x)$ is used as the activation function with  slope coefficient $\alpha$ set to $0.2$. Adam optimizer \cite{AdamOpt} is used as the optimizer to update the weights and biases. The output of each hidden layer is normalized (via batch normalization) before entering the activation function so that the network converges faster both for $G$ and $D$. The input data includes the 10 most recent sensing results of the jammer and the channel labels. Conditional GAN is trained for $3500$ (deep learning training) epochs where the generator network learns the distribution of the input data. The losses of $G$ and $D$ are shown in Fig.~\ref{fig:dgloss}. We run the GAN training over 13500 epochs and observe that the initial fluctuations in $G$ and $D$ losses drop significantly after around 3000 epochs. Therefore, we stopped training the GAN at 3500 epochs.

We assume that during the training of the GAN and the subsequent deep learning used for transmission decision prediction, channel statistics and background transmitter's behavior (an unknown decision function) do not change and therefore the transmitter's classifier (to be inferred by the jammer) does not change, as well. The first part of the jammer's algorithm is data collection and training, and the second part is applying the trained classifier to make attack decisions.
Only in the second part, we need to ensure that the algorithm is fast enough, i.e., it is much shorter than channel coherence time.
To make one decision, the jammer needs to apply a pre-trained deep learning network on a given sample. This process is very fast. The test time per sample is measured as $0.024$ milliseconds (much smaller than the channel coherence time).

In the first part of the algorithm, assuming the resolution of the received signal strength indicator (RSSI) levels is 1 second and $500$ measurements are made (realistically needed to infer the transmitter's classifier), it takes $500$ seconds to collect $500$ RSSI levels without using the GAN. In our tests, it takes $23$ seconds to train the GAN based on $10$ real samples (collected over $10$ seconds) using a GeForce GTX 1080 GPU and generate $500$ synthetic samples from the GAN (this second part is relatively very short).  So, overall it takes $10+23$ seconds to prepare data with the GAN. Hence, it takes either $33$ seconds  with the GAN or $500$ seconds without the GAN before the jammer starts training the other deep neural network for success transmission prediction. As a result, the GAN significantly reduces the data collection time of $J$ before it can start jamming.

For training data augmentation, $J$ collects $10$ samples and then applies the conditional GAN to generate $500$ synthetic samples. The performance results with the GAN are summarized in Table \ref{table:gan}. If $J$ builds its classifier $C_J$ based on $10$ real data samples, the error probabilities are $19.80\%$ for false alarm and $21.41\%$ for misdetection. By using additional $500$ synthetic data samples, the error probabilities drop to $7.62\%$ for false alarm and to $10.71\%$ for misdetection. These errors are much smaller than the errors when only $10$ real data samples are used and are close to the errors when total of $500$ real data samples are used to train the deep neural network.

\begin{table}
\caption{Performance evaluation with and without the GAN.}
\centering
{\small
\begin{tabular}{c|c|c|c}
\# measured & \# synthetic & Misdetection &  False alarm \\
samples & samples & probability &  probability \\ \hline \hline
10 & 0 & 21.41\% & 19.80\%\\ \hline
10 & 500 & 10.71\% & 7.62\% \\ \hline
500 & 0 & 10.90\% & 4.48\%
\end{tabular}
}
\label{table:gan}
\end{table}

\section{Defense Scheme}
\label{sec:defense}

\subsection{Defense against Adversarial Deep Learning-based Jamming}

We design a defense scheme where $T$ changes the labels for some samples such that $J$ cannot build a reliable classifier in an exploratory attack, i.e., $T$ poisons the training process of $J$ by providing wrong training data. This corresponds to a \emph{causative} (or \emph{poisoning}) attack of $T$ back at $J$ as a defense scheme.  Against the jamming attack, $T$ needs to change the label  ``ACK'' (i.e., ``a successful transmission'') to ``No ACK" (i.e., ``no successful transmission''), and vice versa. For that purpose, $T$ flips labels, i.e.,
\begin{itemize}
\item $T$ does not transmit even if channel is predicted as idle, and
\item $T$ transmits even if channel is predicted as busy.
\end{itemize}
It is clear that $T$ wants to limit the extent of defense operations such that the overhead for defense (i.e., the increased classification error for its own transmission decisions) can be minimized. Otherwise, $T$ would start making a large number of transmission errors and could not sustain a good throughput even without any jammer in presence.
Suppose that $T$ decides to flip labels for $p_d$ percentage (\%) of time slots.
To achieve the best effect, $T$ needs to carefully select in which time slots to perform defense operations by examining the output of its deep learning algorithm.
In fact, a deep learning-based classifier provides not only labels, but also a score for classification.
In particular, there is a classification score in [0,1], namely the likelihood of whether a channel is idle.
If this score is less than a threshold, a time slot is classified as idle; otherwise it is classified as busy.
For $T$'s classifier built in Section~\ref{sec:transmitter}, this threshold is $0.25$, which is chosen to minimize $\max\{ e_{MD}, e_{FA}\}$.
Moreover, if this score is far away from the threshold, then such a classification has a high confidence; otherwise the confidence is low.
Therefore, to maximize the impact on $J$, $T$ should perform defense operations in time slots when the scores close to $0$ or $1$ are obtained, since they correspond to time slots when $T$'s transmission decisions are more predictable.
As a consequence of this defense with different $p_d$ values, $J$ builds different classifiers with different hyperparameters (see Table~\ref{table:para}) compared to the previous case of no defense in Section~\ref{sec:jammer}.
Note that when the number of layers is $1$, the deep neural network reduces to a standard neural network with one hidden layer.

\begin{table}
\caption{Optimized hyperparameter values of the jammer under different levels of defense.}
\centering
{\small
\begin{tabular}{c|c|c|c}
$p_d$ & \# hidden & \# neurons & Activation \\
      & layers & per layer & function \\ \hline \hline
0\% (no defense) & 2 & 50 & Tanh \\ \hline
10\% & 1 & 60 & Sigmoid \\ \hline
20\% & 1 & 90 & Sigmoid \\ \hline
30\% & 2 & 50 & Sigmoid \\ \hline
40\% & 2 & 80 & Tanh \\ \hline
50\% & 2 & 20 & Sigmoid
\end{tabular}
}
\label{table:para}
\end{table}

Table~\ref{table:defense} shows the results when $T$ performs different numbers of defense operations. We can see that even when $T$  makes deliberately wrong decisions only over $10\%$ of all time slots (i.e., $p_d = 10\%$), $J$'s error probabilities increase significantly, i.e., misdetection probability increases from $4.18\%$ to $17.53\%$ and false alarm probability increases from $14.53\%$ to $23.68\%$. We also calculate the performance of $T$ when $J$ performs a jamming attack in any time slot when $T$ can have a successful transmission if not jammed. With more defense operations (i.e., when more labels are flipped and therefore $p_d$ is large), $T$ can increase its throughput and success ratio. However, if $T$ takes too many (e.g., $p_d = 30\%$) defense operations, its performance starts dropping as its spectrum sensing decisions become more unreliable and its transmission becomes less likely to succeed even in the absence of jamming. Note that the success ratio is defined as the ratio of the successful transmissions over the total number of transmissions. With the defense mechanism, the transmitter can choose to decrease the total number of transmissions. As a result, the throughput can increase with the decreasing success ratio.

\begin{table*}
\caption{Results for defense scheme against adversarial deep learning-based jamming attack.}
\centering
{\small
\begin{tabular}{c|c|c|c|c}
$p_d$ & \multicolumn{2}{|c|}{Jammer error probabilities} & \multicolumn{2}{|c}{Transmitter performance} \\ \cline{2-5}
 & Misdetection & False alarm & Throughput (packet/slot) & Success ratio \\ \hline \hline
0\% (no defense) & 4.18\% & 14.53\% & 0.050 & 6.25\% \\ \hline
10\% & 17.53\% & 23.68\% & 0.132 & 17.98\% \\ \hline
20\% & 32.80\% & 33.33\% & 0.216 & 31.67\% \\ \hline
30\% & 33.92\% & 38.25\% & 0.194 & 30.41\% \\ \hline
40\% & 35.83\% & 37.31\% & 0.178 & 31.67\% \\ \hline
50\% & 38.97\% & 38.33\% & 0.170 & 32.32\%
\end{tabular}
}
\label{table:defense}
\end{table*}

\subsection{Defense against Sensing-based and Random Jamming}

Next, we apply this defense against sensing-based and random jammers. Table~\ref{table:defenseSB1} shows the results under sensing-based jamming when the threshold $\tau$ is set to $3.4$ (as determined to be the best threshold for $J$ in Section \ref{sec:sensing-based}). $J$'s sensing threshold remains fixed during the defense operation. $J$ continues to jam the signal whenever the transmitter flips its decision from ``do not transmit'' to ``transmit''. As a result, both misdetection and false alarm rates decrease. The throughput of $T$ also decreases due to the additional jamming and the decision made not to transmit when the channel is available. The proposed attack mitigation technique is not successful with sensing-based jamming compared to the jamming based on adversarial deep learning.

\begin{table*}
\caption{Results for defense scheme against sensing-based jamming attack when $\tau$ = $3.4$.}
\centering
{\small
\begin{tabular}{c|c|c|c|c}
$p_d$ & \multicolumn{2}{|c|}{Jammer error probabilities} & \multicolumn{2}{|c}{Transmitter performance} \\ \cline{2-5}
& Misdetection & False alarm & Throughput (packet/slot) & Success ratio \\ \hline \hline
0\% (no defense) & 12.8\% & 12.6\% & 0.140 & 16.99\% \\ \hline
10\% & 11.0\% & 12.2\% & 0.124 & 16.32\% \\ \hline
20\% & 10.8\% & 11.0\% & 0.120 & 16.81\% \\ \hline
30\% & 9.20\% & 10.4\% & 0.100 & 15.72\% \\ \hline
40\% & 8.40\% & 7.8\% & 0.098 & 17.31\% \\ \hline
50\% & 7.40\% & 8.0\% & 0.084 & 16.67\%
\end{tabular}
}
\label{table:defenseSB1}
\end{table*}

Table~\ref{table:defense2} shows the results under random jamming with jammer probability of misdetection and false alarm both set to $50\%$ (i.e., when the jammer transmit probability $p_J$ is $0.5$). We can see that random jamming  is not effective, as well, and results in higher error probabilities compared to jamming based on adversarial deep learning.

\begin{table*}
\caption{Results for defense scheme against random jamming attack when jammer transmit probability is $0.5$.}
\centering
{\small
\begin{tabular}{c|c|c}
$p_d$ & \multicolumn{2}{|c}{Transmitter performance} \\ \cline{2-3}
& Throughput (packet/slot) & Success ratio \\ \hline \hline
0\% (no defense) & 0.383 & 47.88\% \\ \hline
10\% & 0.345 & 46.58\% \\ \hline
20\% & 0.306 & 45.06\% \\ \hline
30\% & 0.268 & 43.24\% \\ \hline
40\% & 0.230 & 41.04\% \\ \hline
50\% & 0.192 & 38.30\%
\end{tabular}
}
\label{table:defense2}
\end{table*}

\begin{figure}
\centering
\includegraphics[width=0.95\columnwidth]{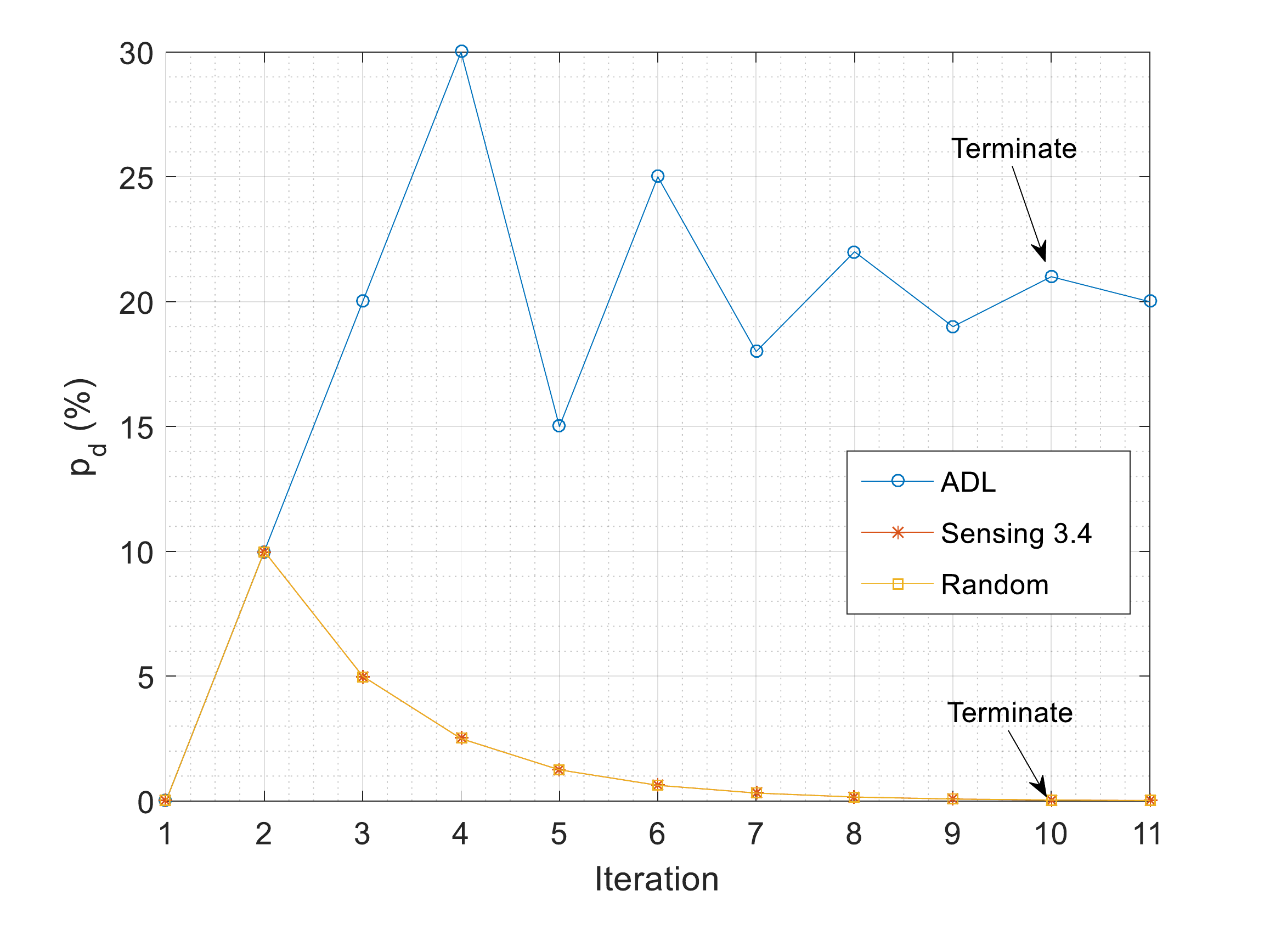}
\caption{The search for the best $p_d$ value in the transmitter's defense scheme.}
\label{fig:pd}
\end{figure}

\subsection{Defense against Unknown Jammer Types}
As expected, the defense mechanism  does not work against random and sensing-based attacks. $T$ may not know what kind of jammer (if any) is launching the attack. Therefore, we introduce an additional step for the transmitter's defense algorithm to start attack mitigation with a fixed level of defense and then gradually increase or decrease the level of defense in response to changes in its throughput that is measured through the received ACK messages.

\begin{figure}
\centering
\includegraphics[width=0.95\columnwidth]{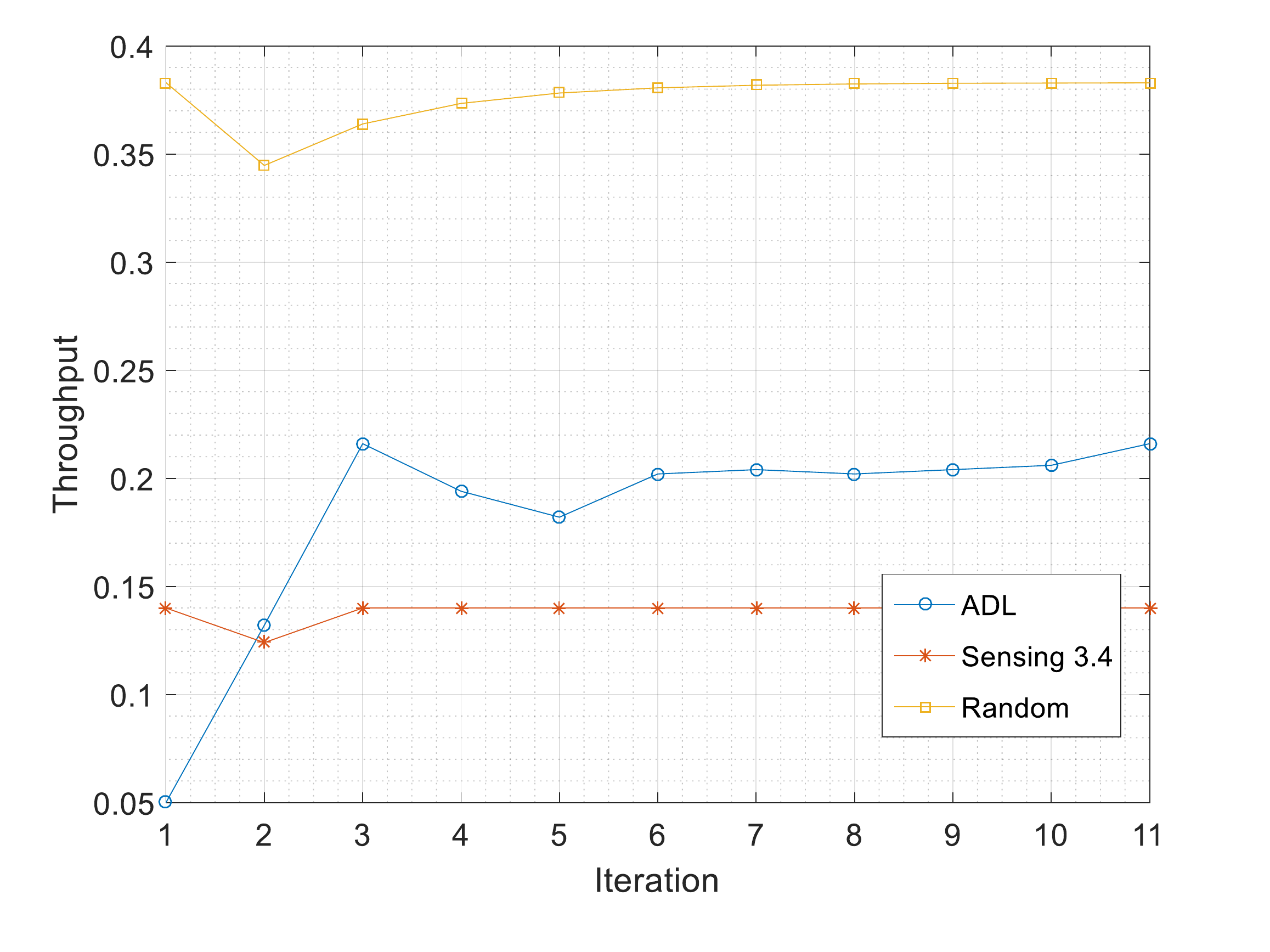}
\caption{The throughput achieved during the search for the best $p_d$ value.}
\label{fig:throughput}
\end{figure}

As we discussed above, if $J$ applies a deep learning based-attack, the optimal defense scheme should have $p_d>0$. If $J$ applies random or sensing-based attack, the optimal defense scheme should have $p_d=0$. In general, $T$ may not know the attack scheme of $J$ and thus we need to design a general approach to search for the best $p_d$ value based on the achieved throughput. A simple approach can be linear search by evaluating a number of points within $[0,100]\%$, e.g., $0\%, 10\%, \cdots, 100\%$, and determining the best one for $p_d$. To have a better granularity, $T$ further searches a number of points around the current best value to find a better $p_d$ value and repeat this process until there is no improvement or the desired granularity is achieved. For example, the best $p_d$ in Table~\ref{table:defense} that maximizes $T$'s throughput is $20\%$. $T$ further searches around $20\%$ with a smaller step size, i.e., $15\%$ or $25\%$, and achieves the throughput of $0.182$ or $0.202$, respectively. So,  $p_d = 20\%$ is still the best. $T$ repeats this process and finally determines that $p_d = 20\%$ is the best choice. For sensing-based jammer (with $\tau=3.4$) and random jammer, $p_d=0\%$ provides the optimal level of defense. This search process is shown in Figs.~\ref{fig:pd} and \ref{fig:throughput} for all jammers, where ``ADL'' refers to  jamming based on adversarial deep learning, ``Sensing 3.4'' refers to sensing-based jamming with  $\tau = 3.4$, and ``Random'' refers to  random jamming. ``Terminate'' points at the iteration where the search of $T$ for the best $p_d$ terminates. $T$ starts using this best $p_d$ value from the next iteration on. Each iteration corresponds to $500$ time slots. In addition to linear search, more advanced searching approaches such as golden section search can also be applied by $T$ to determine $p_d$ with as smaller number of search iterations.  To react to different types of jammers, $T$ starts with some $p_d$ and updates $p_d$ based on its achieved throughput. Fig.~\ref{fig:pd} shows that $p_d$ is reduced over time when $J$ turns out to be a sensing-based or random jammer, whereas $p_d$ is increased to a higher value (beyond which errors in transmit decisions start reducing $T$'s throughput) when $J$ turns out to be an adversarial deep learning-based jammer. Therefore, $T$ can readily adapt when the type of jammer changes over time assuming the convergence time ($10$ iterations) is smaller than the period of changes in the jammer type.

\section{Extension of Network Setting}
\label{sec:ext}
\subsection{Extension to Mobile Setting}

\begin{figure}
\centering
\includegraphics[width=1\columnwidth]{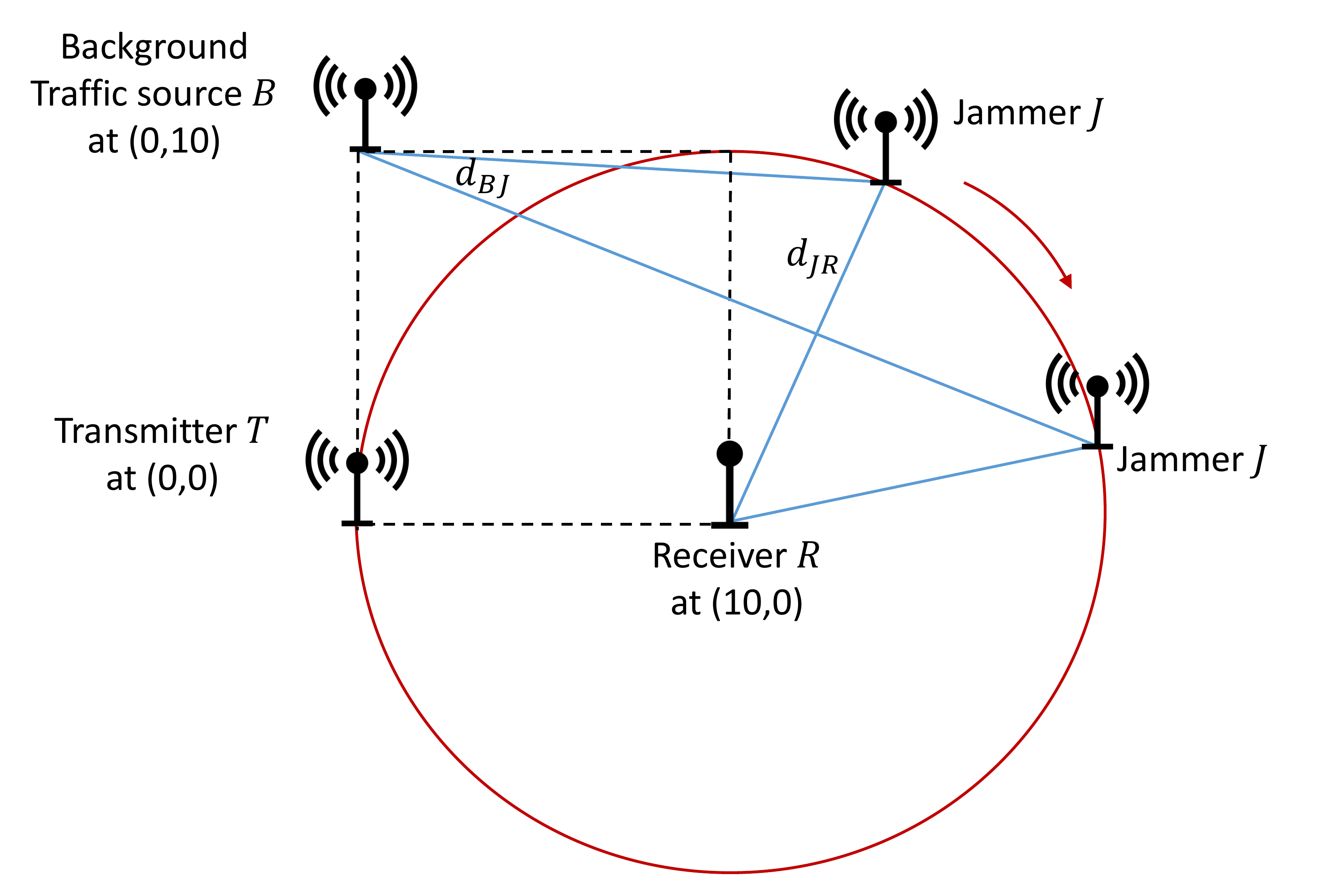}
\caption{The jammer moves along a circle centered at the receiver.}
\label{fig:mobile1}
\end{figure}

The results provided in the previous sections are based on a static network scenario where the positions of $T$, $R$, $B$ and $J$ are fixed at locations $(0,0), (10,0), (0,10)$, and $(10,10)$, respectively. In this section, we evaluate the performance of the proposed system for mobile jammer scenarios. Both transmitter and jammer algorithms can be readily applied in mobile wireless networks. When we considered a static network instance, we did not customize our algorithm to explore the static topology. In the training data, features are derived from sensing results and labels are derived from ACKs, which do not depend on topology. As a result, the same algorithms can be applied when the network is mobile. Note that the distance $d_{BJ}$ between $B$ and $J$ affects $J$'s capability of detection idle/busy channel while the distance $d_{JR}$ between $J$ and $R$ affects the $J$'s capability of jamming. Thus, we fix one of them and study the impact of the other distance. We have the following scenarios.
\begin{itemize}
  \item The distance $d_{JR}$ between $J$ and $R$ is fixed, i.e., $J$ moves along a circle centered at $R$ with radius $10$. In this case, we have $10(\sqrt{2} -1) \le d_{BJ} \le 10(\sqrt{2} +1)$.
  \item The distance $d_{BJ} $ between $B$ and $J$ is fixed, i.e., $J$ moves along a circle centered at $B$ with radius $10$. In this case, we have $10(\sqrt{2} -1) \le d_{JR} \le 10(\sqrt{2} +1)$.
\end{itemize}

\begin{figure}
\centering
\includegraphics[width=0.8\columnwidth]{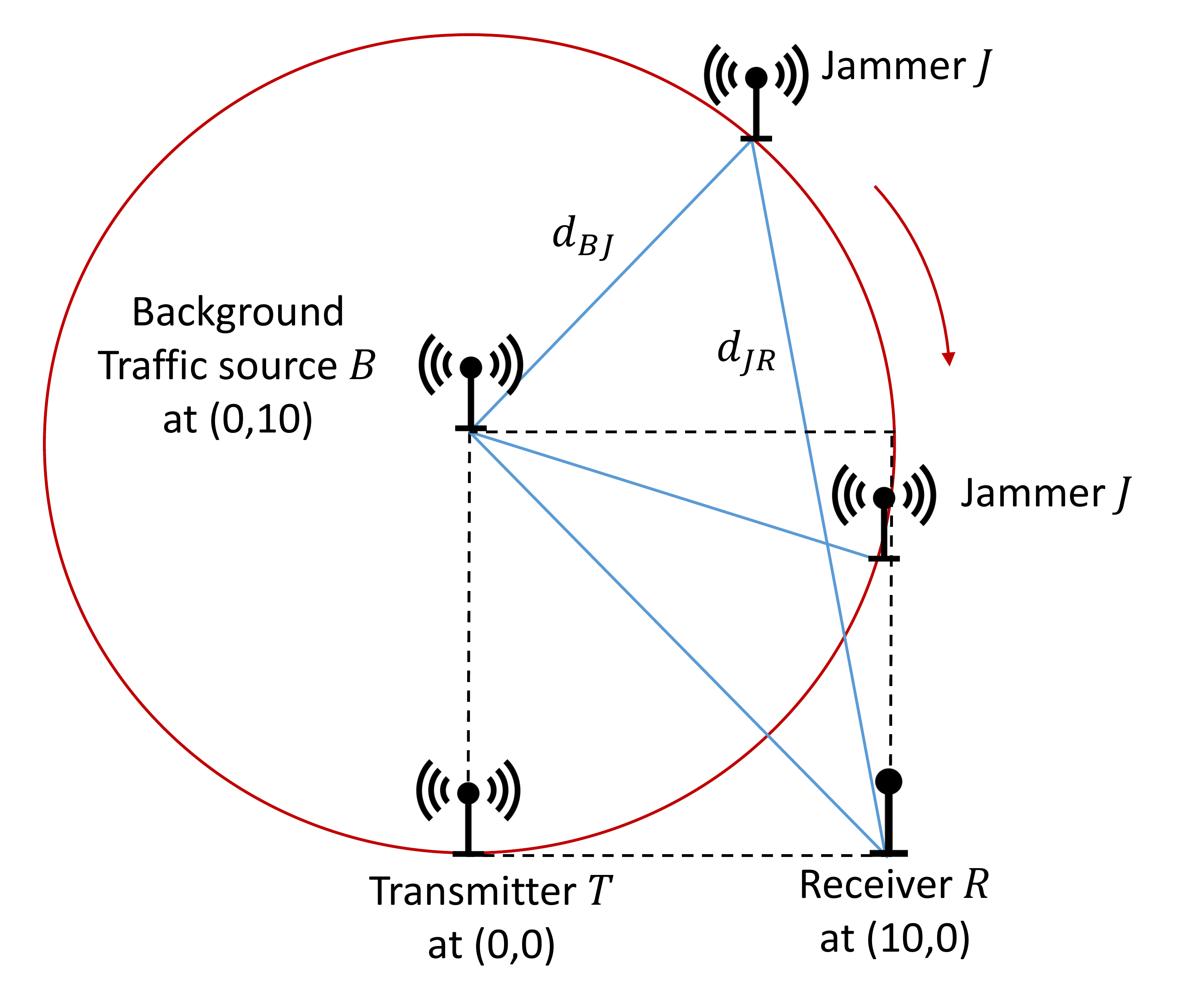}
\caption{The jammer moves along a circle centered at the background traffic source.}
\label{fig:mobile2}
\end{figure}

Fig. \ref{fig:mobile1} shows the scenario when $J$ moves along a circle centered at $R$ and Fig. \ref{fig:mobile2} shows the scenario when $J$ moves along a circle centered at $B$.

For each case, we check five different distance values, i.e., $10(\sqrt{2} -1), 10, 15, 20,$ and $10(\sqrt{2} +1)$. Tables~\ref{table:mobattkcr} and \ref{table:mobattkcbts} show the results when $J$ moves along a circle centered at $R$ and $B$, respectively. Table~\ref{table:mobattkcr} shows that with increasing $d_{BJ}$, it is more challenging for $J$ to learn channel condition and thus its classifier has larger errors. Therefore, $T$ achieves a better performance in terms of throughput and success ratio. Table~\ref{table:mobattkcr} shows that with increasing $d_{JR}$, the impact of jamming is smaller. Therefore, even if the errors to predict channel condition do not change, $T$ achieves a better performance due to less interference.

\begin{table*}
\caption{Results when $J$ moves along a circle centered at $R$.}
\centering
{\small
\begin{tabular}{c|c|c|c|c}
$d_{BJ} $ & \multicolumn{2}{|c|}{Jammer error probabilities} & \multicolumn{2}{|c}{Transmitter performance} \\ \cline{2-5}
& Misdetection & False alarm & Throughput (packet/slot) & Success ratio \\ \hline \hline
4.14 & 1.31\% & 13.68\% & 0.038 & 4.75\% \\ \hline
10 & 4.18\% & 14.53\% & 0.050 & 6.25\% \\ \hline
15 & 5.74\% & 14.53\% & 0.054 & 6.75\% \\ \hline
20 & 13.58\% & 23.93\% & 0.114 & 14.25\% \\ \hline
24.14 & 21.93\% & 23.08\% & 0.182 & 22.75\%
\end{tabular}
}
\label{table:mobattkcr}
\end{table*}

\begin{table*}
\caption{Results when $J$ moves along a circle centered at $B$.}
\centering
{\small
\begin{tabular}{c|c|c|c|c}
$d_{JR}$ & \multicolumn{2}{|c|}{jammer error probabilities} & \multicolumn{2}{|c}{Transmitter performance} \\ \cline{2-5}
 & Misdetection & False alarm & Throughput (packet/slot) & Success ratio \\ \hline \hline
4.14 & 4.18\% & 14.53\% & 0.032 & 4.00\% \\ \hline
10 & 4.18\% & 14.53\% & 0.050 & 6.25\% \\ \hline
15 & 4.18\% & 14.53\% & 0.208 & 26\% \\ \hline
20 & 4.18\% & 14.53\% & 0.436 & 54.5\% \\ \hline
24.14 & 4.18\% & 14.53\% & 0.558 & 69.75\%
\end{tabular}
}
\label{table:mobattkcbts}
\end{table*}

\subsection{Extension of Communication Setting}
The developed solution can be extended for multiple transmitters and receivers, while interference from non-intended transmitters is sensed as the additional interference term by receivers. A transmitter still aims to predict whether the signal strengths at its receiver (if it decides to transmit) will be good or not based on past signal strengths. That is, the only change in the transmitter algorithm is the training data, which includes interference from non-intended transmitters.

We would still consider one jammer. Since each jammer can jam a neighboring area, jammers can be deployed sparsely and each jammer can perform jamming independently. Note that a jammer only needs to predict whether there will be some successful transmissions, i.e., there is no need to figure out corresponding transmitters. Thus, a jammer does not need to build a classifier for each transmitter.
Instead, a jammer aims to predict whether there will be successful transmissions. The only change in the jammer algorithm is the training data, which includes the superimposed signals received from all transmitters and uses the ACKs from all receivers.

\section{Conclusion}
\label{sec:conclusion}

We applied adversarial machine learning to design an intelligent jamming attack on wireless communications and presented a defense scheme against this attack. We considered a wireless communication scenario with one transmitter, one receiver, one jammer, and some background traffic. The transmitter senses the channel and applies a pre-trained machine learning algorithm to detect idle channel instances for transmission. The jammer does not have any knowledge of transmitter's algorithm.
Instead, it senses the channel, detects the transmission feedback (if available), applies a deep learning algorithm to predict a successful transmission, and jams such a transmission. In addition, the jammer uses the deep learning classification scores to control its transmit power subject to an average power constraint.
We showed that this attack is effective in reducing the transmitter's throughput and success ratio compared to random and sensing-based jamming attacks. We also showed that the jammer can effectively apply the GAN to shorten the learning period by augmenting the training data with synthetic samples.
Finally, we designed a defense scheme  for the transmitter that intentionally takes wrong actions in selected time slots to mislead the jammer.
We showed that even a small percentage of wrong actions in systematically selected time slots (based on the transmitter's classification scores) can significantly increase the errors in jammer's decisions and prevent major losses in the performance of the transmitter. This defense scheme does not assume the knowledge of the jammer type and allows the transmitter to adjust its defense level on the fly based on its achieved throughput.

\section*{Acknowledgment}
The authors would like to thank Dr. Kemal Davaslioglu for discussions on the GAN implementation.


\begin{thebibliography}{99}

\bibitem {Clancy2007}
C. Clancy, H. J. Stuntebeck, and T. O'Shea, ``Applications of machine learning to cognitive radio networks,'' \emph{IEEE Wireless Communications}, vol. 14, no. 4, pp.~47--52, 2007.

\bibitem {Thilina2013}
K. Thilina, K. W. Choi, N. Saquib, and E. Hossain, ``Machine Learning Techniques for Cooperative Spectrum Sensing in Cognitive Radio Networks,'' \emph{IEEE Journal on Selected Areas in Communications}, vol. 31, no. 11, pp.~2209--2221, 2013.

\bibitem{XuMobihoc} X. Xu, W. Trappe, Y. Zhang, and T. Wood, ``The Feasibility of Launching and Detecting Jamming Attacks in Wireless Networks,'' \emph{ACM MobiHoc}, 2005.

\bibitem{JammingSurvey}A. Mpitziopoulos, D. Gavalas, C. Konstantopoulos, and G. Pantziou, ``A Survey on Jamming Attacks and Countermeasures in WSNs,'' \emph{IEEE Communications
	Surveys \& Tutorials}, vol. 11, no. 4, pp. 42--56, 2009.

\bibitem{Ateniese}
G. Ateniese, L. Mancini, A. Spognardi, A. Villani, D. Vitali, and G. Felici, ``Hacking Smart Machines with Smarter Ones: How to Extract Meaningful Data from Machine Learning Classifiers,'' \emph{International Journal of Security and Networks}, vol. 10, no. pp.~137--150, 2015.

\bibitem{Tramer}
F. Tramer, F. Zhang, A. Juels, M. Reiter, and T. Ristenpart, ``Stealing Machine Learning Models via Prediction APIs,'' \emph{USENIX Security}, 2016.

\bibitem{Fredrikson}
M. Fredrikson, S. Jha, and T. Ristenpart, ``Model Inversion Attacks that Exploit Confidence Information and Basic Countermeasures,'' \emph{ACM SIGSAC Conference on Computer and Communications Security}, 2015.

\bibitem{Shi17:HST}
Y.~Shi, Y. E.~Sagduyu, and A.~Grushin,
``How to Steal a Machine Learning Classifier with Deep Learning,'' \emph{IEEE Symposium on Technologies for Homeland Security (HST)}, May 2017.

\bibitem{ShiMilcom}
Y. Shi and Y. E Sagduyu, ``Evasion and Causative Attacks with Adversarial Deep Learning," \emph{IEEE Military Communications Conference (MILCOM)}, 2017.

\bibitem {Goodfellow2014}
I. Goodfellow, J. Pouget-Abadie, M. Mirza, B. Xu, D. Warde-Farley, S. Ozair, A. Courville A, and Y. Bengio, ``Generative Adversarial Nets,'' \emph{Advances in Neural Information Processing Systems}, 2014.

\bibitem{Biggio}
B. Biggio, I. Corona, D. Maiorca, B. Nelson, N. Srndic, P. Laskov, G. Giacinto, and F. Roli, ``Evasion Attacks Against Machine Learning at Test Time,'' \emph{ECML PKDD}, 2013.

\bibitem{Kurakin}
A. Kurakin, I. Goodfellow, and S. Bengio, ``Adversarial Examples in the Physical World,'' \emph{arXiv preprint arXiv:1607.02533}, 2016.

\bibitem{Papernot2}
N. Papernot, P. McDaniel,  S. Jha, M. Fredrikson, Z. Celik, and A. Swami, ``The Limitations of Deep Learning in Adversarial Settings,'' \emph{IEEE European Symposium on Security and Privacy}, 2016.

\bibitem{Pi}
L. Pi, Z. Lu, Y. Sagduyu, and S. Chen, ``Defending Active Learning against Adversarial Inputs in Automated Document Classification,'' \emph{IEEE Global Conference on Signal and Information Processing (GlobalSIP)}, 2016.

\bibitem{ShiHST2018}
Y. Shi, Y. E. Sagduyu, K. Davaslioglu, and J. Li, ``Active Deep Learning Attacks under Strict Rate Limitations for Online API Calls," IEEE Symposium on Technologies for Homeland Security (HST), 2018.

\bibitem{Chen2017}
M. Chen, U. Challita, W. Saad, C. Yin, and M. Debbah, ``Machine Learning for Wireless Networks with Artificial Intelligence: A Tutorial on Neural Networks,'' \emph{arXiv preprint arXiv:1710.02913}, 2017.
\bibitem{Alsheikh2014}
M. Alsheikh, S. Lin, D. Niyato, and H. Tan, ``Machine Learning in Wireless Sensor Networks: Algorithms, Strategies, and Applications,'' \emph{IEEE Communications Surveys \& Tutorials}, vol. 16, no. 4, pp.~1996--2018, 2014.

\bibitem{Lee2017}
W. Lee, M. Kim, D. Cho, and R. Schober, ``Deep Sensing: Cooperative Spectrum Sensing Based on Convolutional Neural Networks,'' \emph{arXiv preprint arXiv:1705.08164}, 2017.

\bibitem{Kemal2018}
K. Davaslioglu and Y. E. Sagduyu, ``Generative Adversarial Learning for Spectrum Sensing,'' \emph{IEEE International Conference on Communications (ICC)}, 2018.

\bibitem{DeepOFDM}H. Ye, G. Y. Li, and B.-H. Juang, ``Power of Deep Learning for Channel Estimation and Signal Detection in OFDM Systems,'' \emph{IEEE Wireless Communications Letters}, vol. 7, no. 1, pp.~114--117, 2018.

\bibitem{OShea2016}
T. O'Shea, J. Corgan, and C. Clancy, ``Convolutional radio modulation recognition networks,''  \emph{International Conference on Engineering Applications of Neural Networks}, 2016.

\bibitem{SecurePhy} Y. Zou, J. Zhu, L. Yang, Y. Liang, and Y. Yao ``Securing Physical-Layer Communications for Cognitive Radio Networks,'' \emph{IEEE Communications Magazine}, vol. 53, no. 9, pp.~48--54, 2015.

\bibitem{JamRes} Q. Wang, K. Ren, P. Ning, and S. Hu, ``Jamming-Resistant Multiradio Multichannel Opportunistic Spectrum Access in Cognitive Radio Networks,'' \emph{IEEE Transactions on Vehicular Technology}, vol. 65, no. 10, pp. 8331--8344, 2016.

\bibitem{UserCentric} L. Xiao, J. Liu, Q. Li, N. B. Mandayam, and H. V. Poor, ``User-Centric View of Jamming Games in Cognitive Radio Networks,'' \emph{IEEE Transactions on Information Forensics and Security}, vol. 10, no. 12, pp. 2578--2590, 2015.

\bibitem{GameJam} G. Thamilarasu, and R. Sridhar, ``Game Theoretic Modeling of Jamming Attacks in Ad hoc Networks,'' \emph{IEEE International Conference on Computer Communications and Networks (ICCCN)}, 2009.

\bibitem{EnDep} C. Pielli, F. Chiariotti, N. Laurenti, A. Zanella, and M. Zorzi, ``A Game-theoretic Analysis of Energy-depleting Jamming Attacks,'' \emph{IEEE  International Conference on Computing, Networking and Communications (ICNC)}, 2017.

\bibitem{StochGame} B. Wang, Y. Wu, K. J. R. Liu, and T. C. Clancy, ``An Anti-jamming Stochastic Game for Cognitive Radio Networks,'' \emph{IEEE Journal on Selected Areas in Communications}, vol. 29, no. 4, pp. 877--889, 2011.

\bibitem{TwoDim} G. Han, L. Xiao, and H. V. Poor, ``Two-Dimensional Anti-Jamming Communication based on Deep Reinforcement Learning,'' \emph{IEEE International Conference on Acoustics, Speech and Signal Processing (ICASSP)}, 2017.

\bibitem{IDS} Z. M. Fadlullah, H. Nishiyama, N. Kato, and M. M. Fouda, ``Intrusion Detection System (IDS) for Combating Attacks Against Cognitive Radio Networks,'' \emph{IEEE Network}, vol. 27, no. 3, pp. 51--56, 2013.

\bibitem{EnergyHarvestingCN} D. T. Hoang, D. Niyato, P. Wang, and D. I. Kim, ``Performance Analysis of Wireless Energy Harvesting Cognitive Radio Networks Under Smart Jamming Attacks,'' \emph{IEEE Transactions on Cognitive Communications and Networking}, vol. 1, no. 2, pp. 200--216, 2015.

\bibitem{Yi2018_2}
Y. Shi, T. Erpek, Y. E. Sagduyu, and J. Li, ``Spectrum Data Poisoning with Adversarial Deep Learning,'' \emph{IEEE Military Communications Conference (MILCOM)}, pp. 407--412, 2018.

\bibitem{Yi2018}
Y. Shi, Y. E Sagduyu, T. Erpek, K. Davaslioglu, Z. Lu, and J. Li, ``Adversarial Deep Learning for Cognitive Radio Security: Jamming Attack and Defense Strategies,'' \emph{IEEE International Conference on Communications (ICC) Workshop on Promises and Challenges of Machine Learning in Communication Networks}, 2018.

\bibitem{Sagduyu2011}
Y. E. Sagduyu, R. Berry, and A. Ephremides, ``Jamming Games in Wireless Networks with Incomplete Information,'' \emph{IEEE Communications Magazine}, vol. 49, no. 8, pp. 112-118, 2011.


\bibitem{CNTK}
Microsoft Cognitive Toolkit (CNTK), https://docs.microsoft.com/en-us/cognitive-toolkit

\bibitem{Backprop} D. E. Rumelhart, G. E. Hinton, and R. J. Williams, ``Learning Representations by Back-propagating Errors,'' \emph{Nature}, vol. 323, pp. 533-536, 1986.

\bibitem {Shrivastava2017}
A. Shrivastava, T. Pfister, O. Tuzel, J. Susskind, W. Wang, and R. Webb, ``Learning from Simulated and Unsupervised Images through Adversarial Training,'' \emph{Conference on Computer Vision and Pattern Recognition (CVPR)}, 2017.

\bibitem {YiISSPIT2018}
Y. Shi, Y. E. Sagduyu, K. Davaslioglu, and J. Li, ``Generative Adversarial Networks for Black-Box API Attacks with Limited Training Data," IEEE International Symposium on Signal Processing and Information Technology (ISSPIT), 2018.

\bibitem{bookchapter2018}
Y. Shi, Y. E. Sagduyu, K. Davaslioglu, and R. Levy, ``Vulnerability detection and analysis in adversarial deep learning," in \emph{Guide to Vulnerability Analysis for Computer Networks and Systems: An Artificial Intelligence Approach}. Cham, Switzerland: Springer, 2018.

\bibitem{Mirza2014} M. Mirza and S. Osindero, ``Conditional Generative Adversarial Nets,'' \emph{arXiv preprint arXiv:1411.1784}, 2014.

\bibitem{Tensorflow} M. Abadi, \emph{et al.}, ``TensorFlow: Large-scale machine learning on heterogeneous systems,''  https://tensorflow.org.

\bibitem{AdamOpt} D. P. Kingma, and J. Ba, ``Adam: A Method for Stochastic Optimization,'' \emph{arXiv preprint arXiv:1412.6980}, 2014.

\end{thebibliography}
\end{document}